\documentclass{PoS}
\usepackage{amsmath}

\newcommand{\be}{\begin{equation}}
\newcommand{\ee}{\end{equation}}
\newcommand{\beq}{\begin{eqnarray}}
\newcommand{\eeq}{\end{eqnarray}}

\title{Noncommutative spectral geometry: A guided tour for theoretical
physicists}

\ShortTitle{Noncommutative spectral geometry: A guided tour for
theoretical physicists}

\author{\speaker{Mairi Sakellariadou}%
\\
        Department of Physics, King's College London, University of
  London, Strand WC2R 2LS, London, U.K.\\
        E-mail: \email{Mairi.Sakellariadou@kcl.ac.uk}}

\abstract {We review a gravitational model based on noncommutative
geometry and the spectral action principle. The space-time geometry is
described by the tensor product of a four-dimensional Riemanian
manifold by a discrete noncommutative space consisting of only two
points. With a specific choice of the finite dimensional involutive
algebra, the noncommutative spectral action leads to the standard model
of electroweak and strong interactions minimally coupled to Einstein
and Weyl gravity. We present the main mathematical ingredients of this
model and discuss their physical implications.  We argue that the
doubling of the algebra is intimately related to dissipation and the
gauge field structure. We then show how this noncommutative spectral
geometry model, a purely classical construction, carries implicit in
the doubling of the algebra the seeds of quantization. After a short
review on the phenomenological consequences of this geometric model as
an approach to unification, we discuss some of its cosmological
consequences. In particular, we study deviations of the Friedmann
equation, propagation of gravitational waves, and investigate whether
any of the scalar fields in this model could play the r\^ole of the
inflaton.}

\FullConference{Proceedings of the Corfu Summer Institute 2011 "School
  and Workshops on Elementary Particle Physics and Gravity"\\

                 September 4-18, 2011\\

                 Corfu, Greece}

\begin{document}

\section{Introduction}
One of the main open issues in theoretical high energy physics is the
unification of all forces, including gravity, so that all interactions
correspond to one underlying symmetry.  At low energy scales, one can
consider an effective theory with physics being described by the sum
of the Einstein-Hilbert and the Standard Model action.  However, while
the first part of this action is based upon diffeomorphism invariance,
the second one is based upon internal symmetries of a gauge group.  This
different nature of symmetries for the two parts of the effective
action may be  at the origin to the difficulty of finding a unified theory
of all interactions, including gravity.  As one approaches the Planck
energy scale the quantum nature of space-time reveals itself and this
simplistic effective theory breaks down. Close to the Planck scale
the appropriate formulation of geometry should be within a quantum
framework and the nature of space-time would change in a way so that
one can recover the low energy picture of diffeomorphism and internal
gauge symmetries. A proposal that could lead to a quantum nature of
space-time has been introduced within noncommutative geometry.

In the framework of noncommutative spectral geometry, gravity and the
standard model fields were put together into matter and geometry on a
noncommutative space made from the product of a four-dimensional
standard commutative manifold by a noncommutative internal space.  The
approach is based on a simple idea: using a very simple mathematical
input, namely the choice of a finite dimensional algebra, one can
derive~\cite{ccm} the full complexity of the standard model Lagrangian
coupled to gravity, by employing the formalism of noncommutative
geometry and spectral action.

Noncommutative spectral geometry offers an elegant approach to
unification, based on the symplectic unitary group in Hilbert space,
rather than on finite dimensional Lie groups. The model offers a
unification of internal symmetries with the gravitational ones. All
symmetries arise as automorphisms of the noncommutative algebra of
coordinates on a product geometry.  Due to the lack of a full quantum
gravity theory, which {\sl a priori} should define the geometry of
space-time at Planckian energy scales, we will follow an effective
theory approach and consider the simplest case beyond commutative
spaces. Thus, below but close to the Planck energy scale, space-time
will be considered as the product of a Riemanian spin manifold by a
finite noncommutative space. At higher energy scales space-time
should become noncommutative in a nontrivial way, while at energies
above the Planck scale the whole concept of geometry may altogether
become meaningless. As a next but highly nontrivial step, one should
consider noncommutative spaces whose limit is the almost commutative
space considered here. 

It is worth clarifying that the noncommutative spectral geometry
approach discussed here, goes beyond the noncommutative geometry
notion employed in the literature to implement the fuzziness of
space-time by means of $[{\bf x}^i, {\bf x}^j]=i\theta^{ij}$, where
$\theta^{ij}$ is an anti-symmetric, real, $d\times d$ ($d$ stands for
the dimension of space-time) matrix, and ${\bf x}^i$ denote spatial
coordinates.

In what follows, we briefly present the elements of noncommutative
spectral geometry~\cite{ncg-book1,ncg-book2} as an approach to
unification and highlight the relation between the doubling of the
algebra and the gauge fields~\cite{mag}, an essential element to make
the link with the standard model of particle physics. We then argue
that the doubling of the algebra is related to dissipation, which
incorporates the seeds of quantization~\cite{mag}.  After a short
review on the phenomenological predictions of this purely geometric
approach to the standard model, we discuss some of its cosmological
consequences~\cite{Sakellariadou:2010nr,Nelson:2008uy,Nelson:2009wr,mmm,Nelson:2010rt,Nelson:2010ru}.

\section{Elements of noncommutative spectral geometry}
We consider the geometry of space-time as being described by the
tensor product ${\cal M}\times{\cal F}$ of a four-dimensional smooth
compact~\footnote{The Euclidean space-time manifold is taken to be
compact for simplicity.}  Riemanian manifold ${\cal M}$ by a tiny
discrete finite noncommutative space ${\cal F}$ composed of just two
points. The geometry is thus described by the product of a continuous
geometry for space-time by an internal geometry for the standard model
of particle physics.  The finite geometry ${\cal F}$ will be chosen so
that it is one of the simplest and most natural finite noncommutative
geometries of the right dimension to solve the fermion doubling
problem.

The noncommutative nature of the finite discrete space ${\cal F}$ is
  given by the spectral triple $({\cal A_F, H_F, D_F})$, where all
  ingredients are finite dimensional. In the spectral triple, ${\cal
  A_F}$ is an involution of operators on the finite-dimensional
  Hilbert space ${\cal H_F}$ of Euclidean fermions and ${\cal D_F}$ a
  self-adjoint unbounded operator in ${\cal H_F}$. The operator 
${\cal D_F}$ is such that $J{\cal D_F}
  = \epsilon'{\cal D_F}J$, where $J$ is an anti-linear isometry of the
  finite dimensional Hilbert space, with the properties
\be
J^2=\epsilon~~,~~J\gamma=\epsilon''\gamma J~;\nonumber
\ee
$\gamma$ is the chirality operator and
$\epsilon,\epsilon',\epsilon''\in \{\pm 1\}$.  

Let us discuss the physical reason for introducing the discrete space
${\cal F}$.  There is a distinction between the metric (spectral)
dimension, specified by the behavior of the eigenvalues of the Dirac
operator, and the KO-dimension (K-theoretic dimension), an algebraic
dimension based on K-theory. We first start with the metric dimension.
The relevant Dirac operator for space-time is the ordinary Dirac
operator on a curved space-time, thus the metric dimension is equal to
four. The internal Dirac operator consists of the fermionic mass
matrix, which has a finite number of eigenvalues, and therefore the
internal metric dimension is equal to zero.  Thus, the metric
dimension of the ${\cal M}\times{\cal F}$ geometry is just four, the
same as that of the ordinary space-time manifold. We proceed with the
KO-dimension. There are 8 possible combinations for the numbers
$\epsilon,\epsilon',\epsilon''$, leading to a KO-dimension modulo
8. To resolve the fermion doubling problem, by projecting out the
unphysical degrees of freedom resting in the internal space, the real
structure of the finite geometry ${\cal F}$ turns out to be such that
its KO-dimension is equal to six, leading to
$(\epsilon, \epsilon', \epsilon'')=(1,1,-1)$. Setting the KO-dimension
of the product space $\mathcal{M}\times {\cal F}$ to be $10\sim 2\
{\rm modulo}\ 8$, allows one to impose simultaneously the reality and
Weyl conditions in the Minkowskian continued forms.  Thus, the reason
for introducing ${\cal F}$ is to correct the KO-dimension from four to
ten (modulo 8). In other words, the fermion doubling problem
requires~\cite{ac2006,fdp} crossing the ordinary four-dimensional
continuum by a space of KO-dimension 6.

The spectral geometry is given by the product rules:
\be
{\cal A}=C^\infty({\cal M})\oplus{\cal A_F}\ \ ,\ \ 
{\cal H}=L^2({\cal M},S)\oplus{\cal H_F}\ \ , \ \ 
{\cal D}={\cal D_M}\oplus1+\gamma_5\oplus{\cal D_F}~,
\nonumber
\ee
where $L^2({\cal M}, S)$ is the Hilbert space of $L^2$ spinors and
${\cal D_M}$ is the Dirac operator of the Levi-Civita spin connection
on the four-dimensional manifold ${\cal M}$. The chirality operator is
$\gamma=\gamma_5\oplus\gamma_{\cal F}$ and the anti-unitary operator
on the complex Hilbert space is $J=J_{\cal M}\oplus J_{\cal F}$, with
$J_{\cal M}$ the charge conjugation.
In order to avoid the fermion doubling problem, one must take
\be
J^2_{\cal F}=1\ \ ,\ \ J_{\cal F}{\cal D}_{\cal F}={\cal D}_{\cal F}
J_{\cal F}\ \ ,\ \
J_{\cal F}\gamma_{\cal F}=-\gamma_{\cal F} J_{\cal F}~.
\ee
In what follows we only consider the noncommutative discrete space
${\cal F}$; to simplify the notation we omit the subscript $_{\cal
F}$.
 
The main input of this purely geometric approach to the standard model
of particle physics is the choice of a finite dimensional involutive
algebra~\cite{ncg-book1,ncg-book2}. In the context of left-right
symmetric models, the algebra is a direct sum of the matrix algebras $
M_N(\mathbb{C})$ with $N=1,3$ with two copies $\mathbb{H}_{\rm L},
\mathbb{H}_{\rm R}$ of the algebra of quaternions $\mathbb{H}$, namely
\be {\cal A}_{\rm LR}=\mathbb{C}\oplus \mathbb{H}_{\rm L}\oplus
\mathbb{H}_{\rm R} \oplus M_3(\mathbb{C})~;\nonumber\ee
the {\sl left-right symmetric algebra}.  The
fermions of the standard model can be identified with a basis for a
sum of 3 (i.e. the number of generations, which is considered here to
be equal to 3) copies of the representation of the algebra ${\cal
A}_{\rm LR}$, which is the sum of the irreducible bimodules of odd
spin. The algebra ${\cal A}_{\rm LR}$ admits a natural sub-algebra
$\mathbb{C}\oplus M_3(\mathbb{C})$, corresponding to integer spin.

However, our aim is to construct a model that accounts for massive
neutrinos and neutrino oscillations, thus it cannot be a
left-right symmetric model. We will therefore select a sub-algebra of
the left-right symmetric algebra, which breaks left-right symmetry and
leads to the involutive algebra:
\be
\{(\lambda, q_{\rm L},\lambda, m)\ | \
\lambda\in\mathbb{C}\ ,\ q_{\rm L}\in \mathbb{H}\ ,\ m\in
M_3(\mathbb{C})\}~,\nonumber \ee
isomorphic to $\mathbb{C}\oplus\mathbb{H}\oplus M_3(\mathbb{C})$. The
algebra of quaternions~\footnote{To obtain the Lagrangian of the
standard model of particle physics we assume quaternion linearity.}
$\mathbb{H} \subset M_2(\mathbb{C})$ is
\be
\mathbb{H} = \left\{ 
\left(\begin{array}{cc} \alpha & \beta \\
-\bar\beta & \bar \alpha \end{array}
 \right) \ ; \ \alpha ,\beta \in \mathbb{C} \right\}~.
\nonumber
\ee
Consider a finite dimensional Hilbert space ${\cal H}$ of
dimension $n$, with an anti-unitary operator $J$, such that $J^2=1$.
Noncommutative geometry imposes constraints on the involutive algebras
of operators in the Hilbert space. The involutive algebra ${\cal A}$
must be such that
\be
[a, b^0]=0\ \ ,\  \ \forall a,b\in {\cal A}~,\nonumber
\ee
where $b^0=Jb^\star J^{-1}$ and the representation of ${\cal A}$ and
$J$ in ${\cal H}$ is an irreducible representation.  To get an
irreducible solution, the dimension $n$ must be either $k^2$ or
$2k^2$.  Classifying all irreducible finite noncommutative geometries
of KO-dimension six, it was shown~\cite{cc0706} that only $n=2k^2$ can
avoid fermion doubling.  There are thus six possibilities for the
algebra ${\cal A}$, namely
\be \{M_k(\mathbb{C})\ \ \mbox{or}\ \ M_k(\mathbb{R}) \ \ \mbox{or}\ \
M_a(\mathbb{H}) \}\oplus \{M_k(\mathbb{C})\ \ \mbox{or} \ \
M_k(\mathbb{R})\ \ \mbox{or}\ \ M_a(\mathbb{H})\}~.\nonumber \ee
It turns out that five of these possibilities are ruled
out~\cite{ncg-pub}.  Imposing an anti-linear isometry $I$ such that
$I^2=-1$ in just one of the algebras and letting the other one free,
the algebra ${\cal A}$ must be then the following
one~\cite{Chamseddine:2007ia}:
\be \mathcal{A}=M_{a}(\mathbb{H})\oplus M_{k}(\mathbb{C})\ \ \ \ \mbox
{with}\ \ \ \  k=2a~.\nonumber  \ee
 The choice $k=4$ is the first one that produces the correct number of
$k^2=16$ fermions in each of the three generations~\footnote{If at
the CERN Large Hadron Collider new particles are discovered, one may
be able to include them by considering a higher value for $k$. }; the
number of generations is a physical input.

For commutative geometries, a real variable described by a real-valued
  function on a space is given by the corresponding algebra of
  coordinates, while for noncommutative geometries is represented as
  operators in a fixed Hilbert space.  Since real coordinates are
  represented by self-adjoint operators, all information about space
  is encoded in the algebra of coordinates ${\cal A}$, which is
  related to the gauge group of local gauge transformations. While the
  choice of the algebra constitutes the main input of this model, the
  choice of Hilbert space is irrelevant.

The operator ${\cal D}$ corresponds to the inverse of the Euclidean
propagator of fermions. It is given by the Yukawa coupling matrix
which encodes the masses of the elementary fermions and the
Kobayashi--Maskawa mixing parameters. The commutator $[D,a]$, with
$a\in {\cal A}$, plays the r\^ole of the differential quotient
$da/ds$, with $ds$ the unit of length.  The familiar geodesic formula
\be
d(x,y)={\rm inf}\int_\gamma ds
\ee
(the infimum is taken over all possible paths connecting $x$ to $y$),
which is used in Riemanian geometry to determine the distance
$d(x,y)$ between two points $x$ and $y$, is
replaced  in noncommutative geometry by
\be
d(x,y)={\rm sup}\{|f(x)-f(y)|: f\in {\cal A}, ||[D,f]|| \leq 1\}
\ee
(where $D$ is the inverse of the line element $ds$).  To describe
noncommutative geometry, we will focus on the Dirac operator ${\cal
D}$, instead of the metric tensor $g_{\mu\nu}$ which is used for
spaces with commuting coordinates. The standard model fermions provide
the Hilbert space ${\cal H}$ of a spectral triple for the algebra
${\cal A}$, while the bosons are obtained through inner fluctuations
of the Dirac operator of the product geometry.

Since all experimental data are of a spectral nature, we aim at
extracting information, from our noncommutative geometry construction,
which is of a spectral nature. The spectral action functional in
noncommutative spaces is analogous to the Fourier transform in spaces
for which spatial coordinates commute.  We then apply the spectral
action principle, stating that the bare bosonic~\footnote{The
fermionic term can be included by adding $(1/2)\langle J\psi,{\cal
D}\psi\rangle$, where $J$ is the real structure on the spectral
triple and $\psi$ is a spinor in the Hilbert space of the quarks and
leptons.  } Euclidean action is the trace of the heat kernel
associated with the square of the Dirac operator and is of the form
${\rm Tr}(f({\cal D}/\Lambda))$; $f$ is a cut-off function and
$\Lambda$ fixes the energy scale. This action can be seen {\sl \`a la}
Wilson as the bare action at energy scale $\Lambda$. Thus, following
the Wilsonian approach, one can obtain physical predictions for the
standard model parameters by running them down to low (present) energy
scales through the renormalization group equations. Let us emphasize
that this picture is only valid at high energies (at the scale
$\Lambda$, taken to be the unification scale) and the spectral action
must be considered in the Wilsonian approach, where all coupling
constants are energy dependent and follow the renormalization group
equations.  Since both ${\cal D}$ and $\Lambda$ have physical
dimensions of a mass, there is no absolute scale on which they can be
measured. The r\^ole of the cut-off scale $\Lambda$ is equivalent to
keeping only frequencies smaller than the mass scale $\Lambda$. Note
that ${\rm Tr}(f({\cal D}/\Lambda))$ is the fundamental action
functional that can be used not only at the classical level but also
at the quantum level, after Wick rotation to Euclidean signature.

The formalism of spectral triples favors Euclidean rather than
Lorentzian signature~\footnote{The issue of Euclidean versus
Lorentzian signature is also encountered in the nonperturbative
path-integral approach to quantum gravity.}.  The discussion of
phenomenological/cosmological aspects of the theory relies on a Wick
rotation to imaginary time, into the Lorentzian signature. While
sensible from the phenomenological point of view, there exists as yet
no justification on the level of the underlying theory.

In conclusion, one can obtain the full standard model minimally
coupled to Einstein and Weyl gravity, plus higher order
nonrenormalizable interactions suppressed by powers of the inverse of
the mass scale of the theory, through the action functional~\cite{ccm}
\be {\cal S}={\rm Tr}\left(f\left(\frac{\cal D}{\Lambda}\right)\right)+
\frac{1}{2}\langle
J\psi,D\psi\rangle~~,~~\psi\in{\cal H}^+\, \ee
applied to uni-modular inner fluctuations
\be
{\cal D}\rightarrow {\cal D}_A={\cal D}+A+\epsilon'JAJ^{-1}~;\nonumber
\ee
$A=A^\star$ is a self-adjoint operator of the form
\be
A=\sum_j a_j[{\cal D},b_j]\ \ , \ \ a_j, b_j\in{\cal A}~.\nonumber
\ee
Using heat kernel methods, the trace ${\rm Tr}(f({\cal D}_A/\Lambda))$
can be written in terms of the geometrical Seeley-de Witt coefficients
$a_n$, which are known for any second order elliptic differential
operator, as $\sum_{n=0}^\infty F_{4-n}\Lambda^{4-n}a_n$~, where the
function $F$ is defined such that $F({\cal D}_A^2)=f({\cal D}_A)$.  Thus,
the bosonic part of the spectral action can be expanded in powers of
$\Lambda$ in the form~\cite{ac1996,ac1997}
\begin{equation}
\label{eq:sp-act}
{\rm Tr}\left(f\left(\frac{{\cal D}_A}{\Lambda}\right)\right)\sim
\sum_{k\in {\rm DimSp}} f_{k} \Lambda^k{\int\!\!\!\!\!\!-} |{\cal
  D}_A|^{-k} + f(0) \zeta_{{\cal D}_A(0)}+ {\cal O}(1)~.
\end{equation}
The momenta $f_k$ are defined as $f_k\equiv\int_0^\infty f(u)
u^{k-1}{\rm d}u$ for $k>0$ and $f_0\equiv f(0)$, the noncommutative
integration is defined in terms of residues of zeta functions
$\zeta_{{\cal D}_A} (s) = {\rm Tr}(|{{\cal D}_A}|^{-s})$ at poles of the zeta
function, and the sum is over points in the dimension spectrum
of the spectral triple. 

For the four-dimensional Riemanian geometry, the trace is expressed
perturbatively in terms of the geometrical Seeley-deWitt coefficients
$a_n$, as~\cite{sdw-coeff}:
\be\label{asymp-exp} {\rm Tr}\left(f\left(\frac{{\cal
D}_A}{\Lambda}\right)\right) \sim
2\Lambda^4f_4a_0+2\Lambda^2f_2a_2+f_0a_4+\cdots
+\Lambda^{-2k}f_{-2k}a_{4+2k}+\cdots~.  \ee The smooth even function
$f$, which decays fast at infinity, only enters in the multiplicative
factors:
\beq
f_4&=&\int_0^\infty f(u)u^3 du,\nonumber\\
f_2&=&\int_0^\infty f(u)u du~,\nonumber\\
f_0&=&f(0),\nonumber\\
f_{-2k}&=&(-1)^k\frac{k!}{(2k)!} f^{(2k)}(0)~.
\eeq
Since $f$ is taken as a cut-off function, its Taylor expansion at zero
vanishes, therefore the asymptotic expansion, Eq.~(\ref{asymp-exp}),
reduces to 
\be {\rm Tr}\left(f\left(\frac{{\cal
D}_A}{\Lambda}\right)\right) \sim
2\Lambda^4f_4a_0+2\Lambda^2f_2a_2+f_0a_4~;  \ee
the cut-off function $f$ plays a r\^ole only through its momenta $f_0,
f_2, f_4$, which are three real parameters, related to the coupling
constants at unification, the gravitational constant, and the
cosmological constant, respectively.
In the four-dimensional case, the term in $\Lambda^4$ in the spectral
action, Eq.~(\ref{eq:sp-act}), gives a cosmological term, the term in
$\Lambda^2$ gives the Einstein-Hilbert action functional with the
physical sign for the Euclidean functional integral (provided
$f_2>0$), and the $\Lambda$-independent term yields the Yang-Mills
action for the gauge fields corresponding to the internal degrees of
freedom of the metric. The scale-independent terms in the spectral
action have conformal invariance.

The physical Lagrangian, obtained by applying the spectral action
principle in the product geometry, is entirely determined by the
geometric input, namely the ${\cal M}\times {\cal F}$ space. It
contains, in addition to the full standard model Lagrangian, the
Einstein-Hilbert action with a cosmological term, a topological term
related to the Euler characteristic of the space-time manifold, a
conformal Weyl term and a conformal coupling of the Higgs field to
gravity. The Higgs field is the vector boson of the internal
noncommutative degrees of freedom.  
The bosonic action in Euclidean signature reads~\cite{ccm}
\beq\label{eq:action1} 
{\cal S}^{\rm E} = \int \left(
\frac{1}{2\kappa_0^2} R + \alpha_0
C_{\mu\nu\rho\sigma}C^{\mu\nu\rho\sigma} + \gamma_0 +\tau_0 R^\star
R^\star
\right.  
+ \frac{1}{4}G^i_{\mu\nu}G^{\mu\nu
  i}+\frac{1}{4}F^\alpha_{\mu\nu}F^{\mu\nu\alpha}\nonumber\\ 
+\frac{1}{4}B^{\mu\nu}B_{\mu\nu}
+\frac{1}{2}|D_\mu{\bf H}|^2-\mu_0^2|{\bf H}|^2
\left.
- \xi_0 R|{\bf H}|^2 +\lambda_0|{\bf H}|^4
\right) \sqrt{g} \ d^4 x~, \eeq
where 
\beq\label{bc}
\kappa_0^2&=&\frac{12\pi^2}{96f_2\Lambda^2-f_0\mathfrak{c}}~,\nonumber\\
\alpha_0&=&-\frac{3f_0}{10\pi^2}~,\nonumber\\
\gamma_0&=&\frac{1}{\pi^2}\left(48f_4\Lambda^4-f_2\Lambda^2\mathfrak{c}
+\frac{f_0}{4}\mathfrak{d}\right)~,\nonumber\\
\tau_0&=&\frac{11f_0}{60\pi^2}~,\nonumber\\
\mu_0^2&=&2\Lambda^2\frac{f_2}{f_0}-{\frac{\mathfrak{e}}{\mathfrak{a}}}~,
\nonumber\\
\xi_0&=&\frac{1}{12}~,\nonumber\\
\lambda_0&=&\frac{\pi^2\mathfrak{b}}{2f_0\mathfrak{a}^2}~; \eeq
${\bf H}$ is a rescaling 
\be {\bf H}=(\sqrt{af_0}/\pi)\phi~, \nonumber \ee
of the Higgs field $\phi$ to normalize the kinetic energy, and the
momentum $f_0$ is physically related to the coupling constants at
unification.  

Notice the absence of quadratic terms in the curvature; there is only
the term quadratic in the Weyl curvature and the topological term
$R^\star R^\star$. In a cosmological setting, namely for
Friedmann-Lema\^{i}tre-Robertson-Walker  geometries, the Weyl
term vanishes. Notice also the term that couples gravity with the
standard model, a term which should always be present when one
considers gravity coupled to scalar fields.  It is important to emphasize
that the relations in Eq.~(\ref{bc}) are tied to the scale at which
the expansion is performed.  There is {\sl a priori} no reason for the
constraints to hold at scales below the unification scale $\Lambda$,
since they represent mere boundary conditions~\footnote{One can find
in the literature the unjustified {\sl ansatz} that these boundary
conditions are functions of the energy scale.}.

The geometric parameters $\mathfrak{a}, \mathfrak{b}, \mathfrak{c},
\mathfrak{d}, \mathfrak{e}$ describe the possible choices of Dirac
operators on the finite noncommutative space. These parameters
correspond to the Yukawa parameters of the particle physics model and
the Majorana terms for the right-handed neutrinos. They are given
by~\cite{ccm}
\beq\label{eq:Ys-oth}
 \mathfrak{a}&=&{\rm Tr}
\left( Y^\star_{\left(\uparrow 1\right)} Y_{\left(\uparrow 1\right)} +
Y^\star_{\left(\downarrow 1\right)} Y_{\left(\downarrow
  1\right)}
+ 
3\left( Y^\star_{\left(\uparrow 3\right)} Y_{\left(\uparrow 3\right)}
+ Y^\star_{\left(\downarrow 3\right)} Y_{\left(\downarrow 3\right)}
\right)\right)~,\nonumber\\ \mathfrak{b}&=&{\rm Tr}\left(\left(
Y^\star_{\left(\uparrow 1\right)} Y_{\left(\uparrow 1\right)}\right)^2
+ \left(Y^\star_{\left(\downarrow 1\right)} Y_{\left(\downarrow
  1\right)}\right)^2
+ 
3\left( Y^\star_{\left(\uparrow 3\right)} Y_{\left(\uparrow
  3\right)}\right)^2 + 3 \left(Y^\star_{\left(\downarrow 3\right)}
  Y_{\left(\downarrow 3\right)} \right)^2\right)~,\nonumber\\
  \mathfrak{c}&=&{\rm Tr}\left(Y^\star_R Y_R\right)~,\nonumber\\
  \mathfrak{d}&=&{\rm Tr}\left(\left(Y^\star_R
  Y_R\right)^2\right),\nonumber\\ \mathfrak{e}&=&{\rm
  Tr}\left(Y^\star_R Y_RY^\star_{\left(\uparrow 1\right)}
  Y_{\left(\uparrow 1\right)}\right)~, \eeq
with $Y_{\left(\downarrow 1\right)}, Y_{\left(\uparrow 1\right)},
Y_{\left(\downarrow 3\right)}, Y_{\left(\uparrow 3\right)}$ and $Y_R$
being $(3\times 3)$ matrices, with $Y_R$ symmetric. The $Y$ matrices
are used to classify the action of the Dirac operator and give the
fermion and lepton masses, as well as lepton mixing, in the asymptotic
version of the spectral action.  The Yukawa parameters run with the
renormalization group equations of the particle physics model. 

It is worth noting that since running towards lower energies implies
that nonperturbative effects in the spectral action cannot be any
longer neglected, any results based on the asymptotic expansion and on
renormalization group analysis can only hold for early universe
cosmology.  Hence, the spectral action at the unification scale
$\Lambda$ offers a framework to investigate early universe
cosmological models. For later times, one should consider the full
spectral action, a direction which requires the development of
nontrivial mathematical tools.

\section{Dissipation and the origin of quantization}
The central ingredient of the noncommutative spectral geometry model,
namely the doubling of the algebra acting on the doubled Hilbert
space, is also present in the quantum mechanics formalism of the
Wigner function
\beq \label{W} W(p,x,t) = \frac{1}{2\pi \hbar}\int {\Psi^* \left(x -
\frac{1}{2}y,t\right)\Psi \left(x + \frac{1}{2}y,t\right)
e^{-i\frac{py}{\hbar}}dy} ~, \nonumber \eeq 
and the density matrix
\be\label{AA8} 
W(x_{+},x_{-},t) \equiv \langle x_{+}|\rho (t)|x_{-}\rangle =
\Psi^* (x_{-},t)\Psi (x_{+},t)~. \ee
Here we split the coordinate $x(t)$ of a quantum particle being into
two coordinates $x_+(t)$ and $x_-(t)$:
\be
 x_+(t)=x(t)+\frac{1}{2}y(t)\ \ \ \ \mbox{and} \ \ \ \ 
x_-(t)=x(t)- \frac{1}{2}y(t)~,
\ee
going forward and backward in time, respectively.  

The forward and backward in time evolution of the density matrix
$W(x_{+},x_{-},t)$ is then described by two copies of the
Schr\"odinger equation, as
\be i\hbar {\partial \langle x_+|\rho (t)|x_-\rangle \over \partial t}=
H \langle x_+|\rho (t)|x_- \rangle~, \label{(5a)} \ee
where $H \,=\, H_+ -H_-$ with $H_{\pm}$ the  two Hamiltonian operators.

Thus, the density matrix and the Wigner function require the
introduction of a doubled set of coordinates and of their respective
algebras.  Equation~(\ref{(5a)}) implies that the eigenvalues of $H$
are directly the Bohr transition frequencies
\be
h \nu_{nm}=E_n-E_m~,\nonumber
\ee
which was the first hint towards an explanation of spectroscopic
structure. This can be seen as the connection between noncommutative
algebra, spectroscopic experiments and energy level discretization.

Moreover, the doubling of the algebra is implicit even in the
classical theory when considering the Brownian motion of a particle with
equation of motion
\be
m\ddot{x}(t)+\gamma \dot{x}(t)=f(t)~; \label{(br-mot)}
\ee
$f(t)$ is a random Gaussian distributed force with
\be <f(t)f(t^\prime )>_{\rm noise}=2\,\gamma \,k_BT\; \delta
(t-t^\prime)~. \label{(gaussian)} \ee
Equation~(\ref{(br-mot)}) can be derived~\cite{Blasone:1998xt} from a
Lagrangian in a canonical procedure, using a delta functional
classical constraint representation as a functional integral. By
averaging over the fluctuating force $f$, one gets
\be \label{(25)}
<\delta[m\ddot{x}+\gamma \dot{x}-f]>_{\rm noise}= \int D y
<\exp[{i\over \hbar} \int dt \;L_f(\dot{x},\dot{y},x,y)]>_{\rm
noise}~, \ee
where
\be {\cal L}_f(\dot{x},\dot{y},x,y)= m\dot{x}\dot{y}+ {\gamma \over
  2}(x\dot{y}-y\dot{x})+fy~. \label{(26)} \ee
Hence, the constraint condition at the classical level
introduced a new coordinate $y$, and the standard Euler-Lagrange
equations are obtained:
\be
{d\over dt}{\partial {\cal L}_f\over \partial \dot{y}}=
{\partial {\cal L}_f\over \partial y} ~ ; \ \ \
{d\over dt}{\partial {\cal L}_f\over \partial \dot{x}}=
{\partial {\cal L}_f\over \partial x}~,
\ee
leading to
\be
m\ddot{x}+\gamma \dot{x}=f ~,\ \  \
m\ddot{y}-\gamma \dot{y}=0~. \label{(27)}
\ee
It is worth noting that the Lagrangian system,
Eqs.~(\ref{(26)})-(\ref{(27)}), above was obtained in a completely
classical context~\footnote{Note that $\hbar$ has been introduced for
dimensional reasons.} in order to build a canonical formalism for a
dissipative system. The $x$-system is an {\it open} system; to set up
the canonical formalism it is required to {\it close} it and this is
done by introducing its time-reversed copy, the $y$-system.  The
resulting $\{x-y\}$-system is a closed one.

To highlight~\cite{mag} the relation between the doubling of the
algebra and the gauge field structure let us consider the equation of
the classical one-dimensional damped harmonic oscillator
\be
m \ddot x + \gamma \dot x + k x  = 0~,
\ee
with time independent $m$, $\gamma$ and $k$.  As we have just
discussed, in the canonical formalism for open systems the doubling
of the degrees of freedom is required in such a way as to complement
the given open system with its time-reversed image, and thus obtain a
globally closed system for which the Lagrangian formalism is well
defined. Considering the oscillator in the doubled $y$-coordinate
\be
m \ddot y - \gamma \dot y + k y = 0~,
\ee
and introducing the coordinates 
\be x_{1}(t) = \frac{x(t) + y(t)}{\sqrt{2}}\ \ \ \ \mbox{and}\ \ \ \
x_{2}(t) = \frac{x(t) - y(t)}{\sqrt{2}}~, \ee
the Lagrangian of this closed system takes the form
\beq 
\label{lagr}
{\cal L} &=& {1
  \over 2m} (m{\dot x_1} + {e_1 \over{c}} A_1)^2 - {1 \over 2m}
(m{\dot x_2} + {e_2 \over{c}} A_2)^2 - {e^2\over 2mc^2}({A_1}^2 +
{A_2}^2) - e\Phi \label{2.24i}~, \eeq
where we have introduced the vector
potential 
\be
A_i = \frac{B}{ 2} \epsilon_{ij} x_j\ \ \ \ \mbox{for}\ \ \ \  i,j = 1,2~,
\ee
with 
\be
B \equiv \frac{\gamma c}{e}~,
\ee
and 
\be
{\epsilon}_{ii} = 0\ \ \ \ ,\ \ \ \ {\epsilon}_{12} = -
{\epsilon}_{21} = 1~;
\ee
The Lagrangian Eq.~(\ref{lagr}) describes two particles with opposite
charges $e_1 = - e_2 = e$ in the potential 
\be
\Phi \equiv
\frac{k}{2e}({x_1}^2 - {x_2}^2) \equiv {\Phi}_1 - {\Phi}_2~,
\ee
with $ {\Phi}_i
\equiv (k/2/e){x_i}^{2}$, in the constant magnetic field
$\bf{B}$ defined by $\bf{B}= \bf {\nabla} \times
\bf{A}$.

Thus, the doubled coordinate, e.g., $x_2$ acts as the gauge field
component $A_1$ to which the $x_1$-coordinate is coupled, and {\sl
vice versa}. In conclusion, the energy dissipated by one of the two
systems is gained by the other, implying that the gauge field acts as
the bath or reservoir in which the system is embedded~\cite{mag}.

Following 't~Hooft's conjecture~\cite{'tHooft:1999gk}, stating that
there are classical deterministic models for which loss of information
might lead to a quantum evolution, we argue~\cite{mag} that the
noncommutative spectral geometry classical construction carries
implicit in its feature of the doubling of the algebra the seeds of
quantization.  We will show that the Hamiltonian of a classical damped
harmonic $x$-oscillator and its time-reversed image, the
$y$-oscillator, belongs to the class of Hamiltonians for which this
conjecture was proposed.  

The system's Hamiltonian can be written
as~\cite{Blasone:2000ew,Blasone:1996yh}
\be \label{pqham}
H = \sum_{i=1}^2p_{i}\, f_{i}(q)\,,
\ee
with the functions $f_1, f_2 $ given by
\be
f_1(q)=2\Omega~~~~,~~~~f_2(q)=-2\Gamma~,\ee
where
\be
\Gamma = {\gamma\over 2 m}~~,~~ \Omega = \sqrt{\frac{1}{m}
(k-\frac{\gamma^2}{4m})}~~~~ \mbox{with}~~ k >\frac{\gamma^2}{4m}~.\ee
The nonvanishing Poisson brackets are $\{q_{i},p_i\} =1$.

The Hamiltonian Eq.~(\ref{pqham}) belongs to the class of Hamiltonians
considered by 't~Hooft. The $f_{i}(q)$ are nonsingular functions of
the canonical coordinates $q_{i}$ and the equations for the $q$'s,
namely $\dot{q_{i}} = \{q_{i}, H\} = f_{i}(q)$), are decoupled from
the conjugate momenta $p_i$. In such a case, there is a complete set of
observables which Poisson commute at all times. This implies that the
system admits a deterministic description even when expressed in terms
of operators acting on some functional space of states $|\Psi\rangle$,
such as the Hilbert space. Such a description in terms of operators
and Hilbert space, does not imply {\em per se} quantization of the
system.  Quantization is achieved only as a consequence of
dissipation.

Let us write the Hamiltonian as
\be
H = H_{\rm I} - H_{\rm II}~,
\ee
with
\beq
&&H_{\rm I} = \frac{1}{2 \Omega {\cal C}} (2 \Omega {\cal C}
- \Gamma J_2)^2 ~~,~~
H_{\rm II} = \frac{\Gamma^2}{2 \Omega {\cal C}} J_2^2~,   \label{split}
\eeq
where the Casimir operator ${\cal C}$ and 
the (second) SU(1,1) generator $J_2$ are
\be\label{ca} {\cal C} = \frac{1}{4 \Omega m}\left[\left(p_1^2  - p_2^2\right)+
m^2\Omega^2 \left(x_1^2 -  x_2^2\right)\right]~, \ee
(taken to be positive)
and
\be
J_2 = \frac{m}{2}\left[\left( {\dot x}_1 x_2 - {\dot x}_2
x_1 \right) - {\Gamma} r^2 \right]
~,
\ee
respectively, and $r$ is given by $r^2=x_1^2-x_2^2$. 

Let us then impose the constraint $J_2 |\Psi\rangle = 0$, which
defines physical states and guaranties that $H$ is bounded from below.
This implies
\beq \label{17}
H |\Psi\rangle= H_{\rm I} |\Psi\rangle=  2\Omega {\cal C}|\Psi\rangle
= \left( \frac{1}{2m}p_{r}^{2} + \frac{K}{2}r^{2}\right) |\Psi \rangle \, ,
\eeq
with $K\equiv m \Omega^2$. Hence, $H_{\rm I}$ reduces to the
Hamiltonian for the two-dimensional isotropic (or radial)
harmonic oscillator $\ddot{r} + \Omega^2 r =0 $.

The physical states are invariant under time-reversal and periodical
with period $\tau = 2\pi/\Omega$. 
The generic state $|\Psi(t)\rangle_{H}$ can be written as
\beq|\Psi(t)\rangle_{H} = {\hat{T}}\left[ \exp\left(
  \frac{i}{\hbar}\int_{t_0}^t 2 \Gamma J_2 dt' \right) \right]
|\Psi(t)\rangle_{H_{\rm I}} ~, \eeq
where
${\hat{T}}$ denotes time-ordering and the constant $\hbar$, with
dimension of an action, is introduced for dimensional reasons.
The states
$|\Psi(t)\rangle_{H}$ and $|\Psi(t)\rangle_{H_{\rm I}}$ satisfy the equations:
\beq 
i \hbar \frac{d}{dt} |\Psi(t)\rangle_{H} &=& H \,|\psi(t)\rangle_{H}~,
\nonumber \\ 
\mbox{and}\ \ \ \ i \hbar \frac{d}{dt} |\Psi(t)\rangle_{H_{\rm I}} &= &2 \Omega
{\cal C} |\Psi(t)\rangle_{H_{\rm I}} \, ,
\eeq
respectively. The periodicity of the physical states imply 
\begin{eqnarray}
|\Psi(\tau)\rangle &= &\exp\left( i\varphi -
\frac{i}{\hbar}\int_{0}^{\tau}\langle \Psi(t)| H |\Psi(t) \rangle
dt\right) |\Psi(0)\rangle \nonumber \\ &=& \exp\left(- i2\pi n\right)
| \Psi(0)\rangle \, ,
\label{eqtH}
\end{eqnarray}
or equivalently,
\be 
\frac{ \langle \Psi(\tau)| H |\Psi(\tau) \rangle }{\hbar} \tau -
\varphi = 2\pi n~~,~~ n = 0, 1, 2, \ldots~.\ee
Using $\tau = 2 \pi/\Omega$ and $\varphi = \alpha \pi$, where $\alpha$ is
a real constant, we thus obtain
\beq\label{spectrum}
\langle \Psi_{n}(\tau)| H |\Psi_{n}(\tau) \rangle= \hbar
\Omega \left( n + \frac{\alpha}{2} \right) ~.
\eeq
The index $n$ signals the $n$ dependence of the state and the
corresponding energy.  Equation~(\ref{spectrum}) gives the effective
$n^{\rm th}$ energy level of the system corrected by its interaction with the
environment. In conclusion, the dissipation term $J_2$ of the
Hamiltonian is responsible for the zero point ($n = 0$) energy $E_{0}
=(\hbar/2) \Omega \alpha$, which is the signature of quantization. In
conclusion, the zero point quantum contribution $E_0$ to the
spectrum of physical states signals the underlying dissipative
dynamics.

\section{High Energy Phenomenology of the Noncommutative Spectral Geometry}  
Let us proceed with a short discussion on the phenomenological
consequences~\cite{ccm} of the noncommutative spectral approach to the
standard model, the most successful particle physics model we have at
hand.

As a consequence of the choice $M_2(\mathbb{H})\oplus M_4(\mathbb{C})$
for the algebra ${\cal A}$ of the discrete space ${\cal F}$, the
spectrum of the fermionic particles (the number of states in the
Hilbert space) per family~\footnote{The number of families is a
physical input.} is predicted to be $4^2=16$. Moreover, the selected
(in order to be consistent with the axioms of noncommutative geometry)
algebra leads to the gauge group of the standard model. Thus, the 16
spinors get the correct quantum number with respect to the standard
model gauge group. The gauge bosons of the standard model gauge group
are the inner fluctuations of the metric along continuous
directions. In addition, there is a Higgs doublet corresponding to the
inner fluctuations along the discrete directions.  The spectral action
approach leads to a mass of this Higgs doublet with a negative sign
and a quartic term with a plus sign, implying the existence of a
mechanism of spontaneous breaking of the electroweak symmetry.

Let us assume that the function $f$ is well approximated
by the cut-off function and ignore higher order terms.  Normalization
of the kinetic terms implies
\be
\label{g's}
\frac{g_3^2f_0}{ 2\pi^2}=\frac{1}{4} ~~\mbox{and}~~ g_3^2=g_2^2=
\frac{5}{ 3}g_1^2~, \ee
leading to
\be \sin^2\theta_{\rm W}=\frac{3}{8}~, \ee
a relation which was also found in the context of SU(5) and SO(10)
grand unified theories.  Since the predicted relations,
Eq.~(\ref{g's}b), from the noncommutative spectral geometry are the
ones that hold for all grand unified theories, this implies that the
spectral action holds at unification scale.

Assuming the big desert hypothesis, the running of the couplings
$\alpha_i=g_i^2/(4\pi)$ with $i=1,2,3$, up to one-loop
corrections~\footnote{Only at one-loop order the renormalization group
equations for the coupling constants $g_i$ are uncoupled from the
other standard model parameters.}, is 
\be \beta_i={1\over (4\pi)^2}b_ig_i^3 \ \ \ \ \mbox{with}\ \ \ \ 
b=\left({41\over 5},-{19\over 6}, -7\right)~.  \ee
Performing one-loop renormalization group analysis for the running of
the gauge couplings and the Newton constant, it was shown~\cite{ccm}
that these do not meet at a point, the error being within just few
percent. The fact that the predicted unification of the coupling
constants does not hold exactly, implies that the big desert
hypothesis is only approximately valid and new physics are expected
between unification and present energy scales. In terms of our
assumption for the cut-off function, the lack of a unique unification
energy implies that even though the function $f$ can be approximated
by the cut-off function there exist small deviations.

The noncommutative spectral geometry model predicts also the existence
of a see-saw mechanism for neutrino masses with large right-handed
neutrino mass of the order of $\Lambda$.  Moreover, it predicts the
constraint:
\be
\sum_\sigma(y_\nu^\sigma)^2+(y_e^\sigma)^2+3(y_u^\sigma)^2+
+3(y_d^\sigma)^2=4g^2~, \ee
on the Yukawa couplings $y^\sigma$ with $\sigma=1,2,3$, at unification
scale. 

The mass of the top quark is given from
\be
m_{\rm top}={1\over\sqrt 2} u k^t~,
\ee
with $u=2M/g$ the vacuum expectation value of the Higgs field and
$k^t$ the top quark Yukawa coupling. We assume that at unification
scale of $\sim 1.1\times 10^{17}$ GeV the value of $g$ is $\sim 0.517$
and the $\tau$ neutrino Yukawa coupling can be neglected.  Then using
the renormalization group equations, the model predicts a top quark
mass of $\sim 179$ GeV, compatible with the experimental value.

In the spectral action, the Higgs coupling is proportional to the
gauge couplings, which restricts the mass of the Higgs.  Using the
cut-off function, this model predicts a heavy Higgs mass. In zeroth
order approximation, it predicts a mass of the Higgs boson
approximately equal to $170 ~{\rm GeV}$, which is ruled out by current
experimental data.  However, this answer is very sensitive to the
value of the unification scale, as well as to deviations of the
spectral function from the cut-off function we have used.  The actual
value of the Higgs mass will be determined by considering higher order
corrections and incorporating them to the renormalization group
equations.  Nevertheless, it is quite encouraging that this purely
geometric approach to the standard model predicted the right order of
magnitude for the Higgs mass.  Given that this noncommutative spectral
geometry model has to be seen as an effective theory, this result is
quite remarkable.

Since the predicted top quark mass is consistent with experimental
data while the predicted Higgs mass is ruled out, one may deduce that
the top quark mass is less sensitive to the ambiguities of the
unification scale than the Higgs mass.  This conclusion may be
understood in the following way. We have splitted the action
functional into the bosonic and the fermionic parts.  The bosonic
action has been then determined by an infinite expansion assuming
convergence of higher order terms. Thus, while for the bosonic part we
have relied on the first terms of the expansion in inverse powers of
the cut-off scale, the fermionic part being much simpler did not
require such an assumption.

Considering an energy scale $\Lambda\sim 1.1\times 10^{17}\ {\rm
  GeV}$, the standard form of the gravitational action and the
  experimental value of Newton's constant at ordinary scales imply
  $\kappa_0^{-1}\sim 2.43\times 10^{18}\ {\rm GeV}$.

Let us also note that this approach to unification does not provide
any explanation of the number of generations, nor leads to constraints
on the values of the Yukawa couplings.

Finally, the parameter $\Lambda$ which has been introduced as a free
 parameter in the spectral action, can be seen as the vacumm
 expectation value of a dynamical (dilaton) field. Such a filed may
 lead to cosmological consequences and it is worth examining whether
 it could play the r\^ole of the inflaton field.

\section{Cosmological consequences}
The noncommutative spectral action lives by construction at high
energy scales, thus providing a natural framework to address early
universe cosmology.  Investigating the cosmological consequences of
the model, one can test its validity and/or constrain its
parameters. In what follows, we review some cosmological aspects of
this purely geometric approach to the standard model.  Let us first
specify the notation and conventions we use.  The signature is taken
$(-,+,+,+)$ and the Ricci tensor is defined as $R_{\mu\nu} =
R^\rho\phantom{}_{\mu\nu\rho}$, with
$R_{\mu\nu\rho}\phantom{}^\sigma\omega_\sigma = \big[
\bigtriangledown_\mu , \bigtriangledown_\nu \big] \omega_\rho$.

The Lorentzian version of the gravitational part of the asymptotic
formula for the bosonic sector of the noncommutative geometry spectral
action, including the coupling between the Higgs field and the Ricci
curvature scalar, reads~\cite{ccm}
\be\label{eq:1.5} {\cal S}_{\rm grav}^{\rm L} = \int \left(
\frac{1}{2\kappa_0^2} R + \alpha_0
C_{\mu\nu\rho\sigma}C^{\mu\nu\rho\sigma} + \tau_0 R^\star
R^\star\right.  
 -\left.  \xi_0 R|{\bf H}|^2 \right)
\sqrt{-g} \ d^4 x~,\ee
leading to he equations of motion~\cite{Nelson:2008uy}
\be\label{eq:EoM2} R^{\mu\nu} - \frac{1}{2}g^{\mu\nu} R +
\frac{1}{\beta^2} \delta_{\rm cc}\left[
  2C^{\mu\lambda\nu\kappa}_{;\lambda ; \kappa} +
  C^{\mu\lambda\nu\kappa}R_{\lambda \kappa}\right]= 
\kappa_0^2 \delta_{\rm cc}T^{\mu\nu}_{\rm matter}~, \ee
where
\be
\beta^2 \equiv -\frac{1}{4\kappa_0^2 \alpha_0}
\ \ \ \ \mbox{and}\ \ \ \
\delta_{\rm cc}\equiv[1-2\kappa_0^2\xi_0{\bf H}^2]^{-1}~.
\ee
Let us first study the low energy regime and then proceed with the
high energy regime.  Depending on whether we are in the former or the
latter one, will specify whether or not the coupling between the Higgs field
and the background geometry can be neglected. 

\subsection{Low energy regime}
In the low energy weak curvature regime, the nonminimal coupling term
between the background geometry and the Higgs field can be neglected,
implying $\delta_{\rm cc}=1$.  For a
Friedmann-Lema\^{i}tre-Robertson-Walker space-time, the Weyl tensor
vanishes, hence the noncommutative spectral geometry corrections to
the Einstein equation vanish~\cite{Nelson:2008uy}.  Thus for such a
background, the constraint~\cite{Stelle} $\beta_{R^2} \geq 3.2\times
10^{-9} {\rm m}^{-1}$, imposed on {\sl ad hoc} curvature squared terms
(of different form but of the same order) does not necessarily hold
within the noncommutative spectral action context.  It is however
important to constrain $\beta$ since a lower limit to $\beta$ can be
equivalently seen as an upper limit to the moment $f_0$ of the cut-off
function used to define the spectral action. Since $f_0$ can be used
to specify the initial conditions on the gauge couplings, a constraint
on $\beta$ corresponds to a restriction on the particle physics model
at unification scale.

Let us briefly summarize how one can constrain
$\beta$~\cite{Nelson:2010rt,Nelson:2010ru} within the noncommutative
spectral geometry model.  Consider linear perturbations around a
Minkowski background metric in the synchronous gauge. The perturbed
metric reads
\be g_{\mu\nu} = {\rm diag} \left( \{a(t)\}^2 \left[
  -1,\left(\delta_{ij} + h_{ij}\left(x\right)\right) \right]\right)~,
\ee
with $a(t)$ the cosmological scale factor. Since we only consider a
flat background, $a(t)=1$ and $\dot a\equiv da/dt=0$. The remaining
gauge freedom can be completely fixed by setting ${\bf \nabla}_i
h^{ij}=0$.

The linearized equations of motion derived from the noncommutative
spectral action for such perturbations are
\be\label{eq:1} \left( \Box - \beta^2 \right) \Box h^{\mu\nu} =
\beta^2 \frac{16\pi G}{c^4} T^{\mu\nu}_{\rm matter}~, \ee 
where $T^{\mu\nu}_{\rm matter}$ is taken to lowest order in
$h^{\mu\nu}$. It is thus independent of $h^{\mu\nu}$
and satisfies the conservation equations
\be
\frac{\partial}{\partial x^\mu} T^\mu_{\ \nu}=0~.
\ee
Since $\beta$ plays the r\^ole of a mas, it has to be real and
positive, implying $\alpha_0 <0$.  For $\alpha_0>0$ the gravitational
waves evolve according to a Klein-Gordon like equation with a
tachyonic mass, and hence the background, which has been considered to
be a Minkowski space, is unstable.  In conclusion, we must restrict to
$\alpha_0<0$ for Minkowski space to be a (stable) vacuum of the
theory.

Let us study the energy lost to gravitational radiation by orbiting
binaries. In the far field limit, $|{\bf r}| \approx |{\bf r} - {\bf
r}'|$ (where ${\bf r}$ and ${\bf r}'$ stand for the locations of the
observer and emitter, respectively), the spatial components of the
general first order solution for a perturbation against a Minkowski
background are given in terms of the quadrupole moment,
\be
D^{ik}\left(t\right) \equiv \frac{3}{c^2}\int  
x^i x^k T^{00}({\bf r},t) \ d{\bf r}~.
\ee
as
\be\label{eq:4} h^{ik}\left( {\bf r},t\right) \approx \frac{2G
  \beta}{3c^4} \int_{-\infty}^{t-\frac{1}{c}|{\bf r}|} \frac{d
  t'}{\sqrt{c^2\left( t-t'\right)^2 - |{\bf r}|^2} } 
          {\cal J}_1 \left( \beta\sqrt{c^2\left( t-t'\right)^2 - |{\bf
              r}|^2}\right) \ddot{D}^{ik}\left(t'\right)~, \ee
where ${\cal J}_1$ is a Bessel function of the first kind, in terms of
the quadrupole moment.

While in the $\beta\rightarrow \infty$ limit the theory reduces to
that of General Relativity and the familiar results for a massless
graviton are recovered, for finite $\beta$ gravitational radiation
contains both massive and massless modes, both of which are sourced
from the quadrupole moment of the system.

In the far field limit, the rate of energy loss for a binary pair of
masses $m_1, m_2$ in a circular (for simplicity) orbit in the
$(xy)$-plane, reads
\be\label{eq:energy} -\frac{{\rm d} {\cal E}}{{\rm d}t} \approx
\frac{c^2}{20G} |{\bf r}|^2 \dot{h}_{ij} \dot{h}^{ij}~,  \ee
with
\beq
&&\dot{h}^{ij}\dot{h}_{ij}= \frac{128\mu^2|\rho|^4 \omega^6 G^2
  \beta^2}{c^8}\left[ f_{\rm c}^2\left(\beta|{\bf
    r}|,\frac{2\omega}{\beta c}\right) + f_{\rm s}^2\left(\beta|{\bf
    r}|,\frac{2\omega}{\beta c}\right)\right]~, \eeq
and the definitions
\beq\label{eq:f1}
\label{eq:f2}
f_{\rm c}\left( x,z\right) &\equiv&
\int_0^\infty \frac{d s}{\sqrt{s^2 + x^2}} {\cal
  J}_1\left(s\right) \cos \left(z\sqrt{ s^2 + x^2} \right)~;\nonumber\\
 f_{\rm s}\left( x,z\right) &\equiv& \int_0^\infty
\frac{d s}{\sqrt{s^2 + x^2}} {\cal J}_1\left(s\right) \sin
\left(z\sqrt{ s^2 + x^2} \right)~,\nonumber
\eeq
The orbital frequency $\omega$ is constant and given by
\be \omega = |\rho|^{-3/2} \sqrt{ G\left( m_1 + m_2\right)}~,  \ee
with $|\rho|$ the magnitude of the separation vector between the two
bodies.

The integrals in Eq.~(\ref{eq:f1}) exhibit a strong resonance behavior
at $z=1$, which corresponds to the critical frequency
\be
\label{critical}
2\omega_{\rm c} =\beta c~,
\ee
around which strong deviations from the familiar results of General
Relativity are expected. This critical (maximum) frequency comes from
the natural length scale (given by $\beta^{-1}$), at which
noncommutative geometry effects become dominant.  For $\omega<
\omega_{\rm c}$, the $\beta \rightarrow \infty$ limit reproduces the
General Relativity result, as it should. Since this is not the case if
$\omega>\omega_{\rm c}$, we conclude that the critical frequency is
the maximum one. Any deviation from the standard result is suppressed
by the distance to the source, at least for orbital frequencies small
compared to $\beta c$.

The form of the gravitational radiation from binary systems can be 
used to constrain $\beta$. For circular
binary orbits we only need to know the orbital frequency and the
distance to the binary system.   The parameter $\beta$ is then
constrained by requiring the magnitude of deviations
from General Relativity to be less than the uncertainty.  
Thus, for $\omega < \omega_{\rm c}$ we get a lower limit on $\beta$~\cite{Nelson:2010ru}:
\be 
\label{constr-beta}
\beta > 7.55\times 10^{-13}~{\rm m}^{-1}~.
\ee
Due to the large distances to these binary systems, the constraint is
almost exactly due to $\beta > 2\omega /c$. Thus, the strongest
constraint comes from systems with high orbital frequencies.  Future
observations of rapidly orbiting binaries, relatively close to the
Earth, could thus improve this constraint by many orders of magnitude.

\vskip1.truecm

Let us go back to the background equations.  In order for the
corrections to Einstein's equations to be apparent at the level 
of the background, we need  to consider anisotropic models. 
We will thus derive the modified Friedmann
equation for the Bianchi type-V model, for which the space-time
metric in Cartesian coordinates reads
\be g_{\mu\nu} = {\rm diag} \left[ -1,\{a_1(t)\}^2e^{-2nz} ,
  \{a_2(t)\}^2e^{-2nz}, \{a_3(t)\}^2 \right]~, \ee
where $a(t)$, $b(t)$ and $c(t)$ are, in general, arbitrary functions
and $n$ is an integer. 

Defining
$A_i\left(t\right) = {\rm ln} a_i\left(t\right)$  with $i=1,2,3$,
the modified Friedmann equation reads~\cite{Nelson:2008uy}:
\beq\label{eq:Friedmann_BV} \kappa_0^2 T_{00}=&&\nonumber\\
 - \dot{A}_3\left(
\dot{A}_1+\dot{A}_2\right) -n^2 e^{-2A_3} \left( \dot{A}_1
\dot{A}_2-3\right)
&& \nonumber \\
 +\frac{8\alpha_0\kappa_0^2 n^2}{3} e^{-2A_3} \left[
  5\left(\dot{A}_1\right)^2 + 5\left(\dot{A}_2\right)^2 -
  \left(\dot{A}_3\right)^2\right.
&&\nonumber\\
\left. - \dot{A}_1\dot{A}_2 - \dot{A}_2\dot{A}_3
  -\dot{A}_3\dot{A}_1 - \ddot{A}_1 - \ddot{A}_2 - \ddot{A}_3 + 3
  \right]
- \frac{4\alpha_0\kappa_0^2}{3} \sum_i \Biggl\{
\dot{A}_1\dot{A}_2\dot{A}_3 \dot{A}_i
&&\nonumber\\
 + \dot{A}_i \dot{A}_{i+1} \left(
\left( \dot{A}_i - \dot{A}_{i+1}\right)^2 -
\dot{A}_i\dot{A}_{i+1}\right)
 + \left( \ddot{A}_i + \left( \dot{A}_i\right)^2\right)\left[
  -\ddot{A}_i - \left( \dot{A}_i\right)^2 + \frac{1}{2}\left(
  \ddot{A}_{i+1} + \ddot{A}_{i+2} \right)\right.&&\nonumber\\
\left. + \frac{1}{2}\left(
  \left(\dot{A}_{i+1}\right)^2 + \left( \dot{A}_{i+2}\right)^2 \right)
  \right]
%
+ \left[ \dddot{A}_i + 3 \dot{A}_i \ddot{A}_i -\left( \ddot{A}_i +
  \left( \dot{A}_i\right)^2 \right)\left( \dot{A}_i - \dot{A}_{i+1} -
  \dot{A}_{i+2} \right)\right]&&\nonumber\\
\times\left[ 2\dot{A}_i
  -\dot{A}_{i+1}-\dot{A}_{i+2} \right]\Biggr\}~. \eeq
The correction terms in this modified Friedmann equation come in two
types. The first one contains terms which are fourth order in time
derivatives, and thus for the slowly varying functions, usually used
in cosmology, they can be taken to be small corrections.  The second
one occurs at the same order as the standard Einstein-Hilbert terms,
however being proportional to $n^2$, it vanishes for homogeneous
versions of Bianchi type-V. Thus, although anisotropic cosmologies do
contain corrections due to the additional noncommutative spectral
geometry terms in the action, they are typically of higher order.
Inhomogeneous models do contain correction terms that appear on the
same footing as the original (commutative) terms. In conclusion, the
corrections to Einstein's equations are present only in inhomogeneous
and anisotropic space-times.

\subsection{High energy regime}

At energies approaching the Higgs scale, the nonminimal coupling of
the Higgs field to the curvature can no longer be neglected, leading
to corrections even for background cosmologies. To understand the
effects of these corrections let us neglect the conformal term in
Eq.~(\ref{eq:EoM2}), i.e. set $\beta=0$. The equations of motion then
become~\cite{Nelson:2008uy}
\be R^{\mu\nu} - \frac{1}{2}g^{\mu\nu}R =
\kappa_0^2\left[\frac{1}{1-\kappa_0^2 |{\bf H}|^2/6}\right] T^{\mu\nu}_{\rm
  matter}~. \ee 
Hence, $|{\bf H}|$ leads to an effective gravitational constant.

Alternatively, consider the effect of this term on the equations of
motion for the Higgs field in a constant gravitational field.  For
constant curvature, the self interaction of the Higgs field is
increased, since
\be -\mu_0 |{\bf H}|^2 \rightarrow -\left( \mu_0 + \frac{R}{12}\right)
|{\bf H}|^2~.  \ee

\vskip1.truecm 

The nonminimal coupling between the Higgs field and the Ricci
curvature may turn out to be particularly useful in early universe
cosmology~\cite{Nelson:2009wr,mmm}.  Such a coupling has been
introduced {\sl ad hoc} in the literature, in an attempt to drive
inflation through the Higgs field.  However, the coupling constant
between the scalar field and the background geometry is not a free
parameter which could be tuned to achieve a successful inflationary
scenario, it should be instead dictated by the underlying theory.

In a Friedmann-Lema\^{i}tre-Robertson-Walker metric, the Weyl tensor
  vanishes, while the nondynamical term is also neglected. Thus the
  Gravity-Higgs sector of the asymptotic expansion of the spectral
  action, in Lorentzian signature reads
\be
{\cal S}^{\rm
  L}_{\rm GH}=\int\Big[\frac{1-2\kappa_0^2\xi_0
    H^2}{2\kappa_0^2}R 
-\frac{1}{2}(\nabla  H)^2- V(H)\Big] \sqrt{-g}\  d^4x~,
\ee
where 
\be\label{higgs-pot}
V(H)=\lambda_0H^4-\mu_0^2H^2~,
\ee
with $\mu_0$ and $\lambda_0$ subject to radiative corrections as
functions of energy.  For large enough values of the Higgs field, the
renormalized value of these parameters must be calculated, while the
running of the top Yukawa coupling and the gauge couplings must be
evolved simultaneously.

At high energies the mass term is sub-dominant and can be neglected,
thus only the first term in Eq.~(\ref{higgs-pot}) survives.  For each
value of the top quark mass, there is a value of the Higgs mass where
the effective potential is on the verge of developing a metastable
minimum at large values of the Higgs field and the Higgs potential is
locally flattened~\cite{mmm}.  Since the region where the potential
becomes flat is narrow, slow-roll must be very slow, in order to
provide a sufficiently long period of quasi-exponential expansion and
thus solve the shortcomings of the standard hot big bang cosmological
model. If the inflaton field is also going to source the initial
density fluctuations then besides the constraints on the slow-roll
parameters $\epsilon, \eta$ to get sufficient number of e-foldings,
one should also check whether the amplitude of density perturbations
$\Delta_\mathcal{R}^2$ in the spectrum of the cosmic microwave
background temperature anisotropies is in agreement with
measurements. Inflation predicts that at horizon crossing (denoted by
stars), the amplitude of density perturbations is related to the
inflaton potential $V$ through
\be \left(\frac{V_*}{\epsilon_*}\right)^{\frac14}
=2\sqrt{3\pi}\ m_\text{Pl}\ \Delta_\mathcal{R}^\frac12~,
\ee
where $\epsilon_*\leq1$ and $m_{\rm Pl}$ stands for the Planck mass.  Its
value, as measured by WMAP7~\cite{Larson:2010gs}, requires
\be \left(\frac{V_*}{\epsilon_*}\right)^{\frac14}
=(2.75\pm0.30)\times 10^{-2}\ m_\text{Pl}\,~.\label{eq:cobe} \ee
We can then calculate the renormalization of the Higgs self-coupling
and construct an effective potential which fits the
renormalization group improved potential around the flat region.  By
doing this calculation up to two-loops we have found~\cite{mmm} that
around the plateau (the minimum of the potential), there is a very
good analytic fit to the Higgs potential, which takes the form
\be
V^\text{eff}=\lambda_0^\text{eff}(H)H^4=[a\ln^2(b\kappa H)+c] H^4~,
\ee
where the parameters $a, b$ are found to relate to the low energy
values of top quark mass $m_{\rm t}$ as
\begin{align}
a(m_\text{t})&=4.04704\times10^{-3}-4.41909\times10^{-5}
\left(\frac{m_\text{t}}{\text{GeV}}\right)
+1.24732\times10^{-7}\left(\frac{m_\text{t}}{\text{GeV}}\right)^2~,
\nonumber\\ 
b(m_\text{t})&=\exp{\left[-0.979261
\left(\frac{m_\text{t}}{\text{GeV}}-172.051\right)\right]}~.
\end{align}
The third parameter, $c=c(m_\text{t},m_\phi)$, encodes the appearance
of an extremum and depends on the values for top quark mass and Higgs
mass.  An extremum occurs if and only if $c/a\leq 1/16$, the
saturation of the bound corresponding to a perfectly flat region.  It
is convenient to write $c=[(1+\delta)/16]a$, where $\delta=0$
saturates the bound below which a local minimum is formed.

These results have been obtained for the case of minimal coupling,
whereas in noncommutative spectral action there is a small nonminimal
coupling, $\xi_0=1/12$.  The corrections due to conformal coupling to
the potential imply that flatness does not occur at $\delta=0$ anymore
but for fixed values of $\delta$ depending on the value of the top
quark mass. More precisely, for inflation to occur via this mechanism,
the top quark mass fixes the Higgs mass extremely accurately.
Scanning through the parameter space it emerges that sufficient
$e$-folds are indeed generated provided there is a suitably tuned
relationship between the top quark mass and the Higgs mass. In
conclusion, while the Higgs potential can lead to the slow-roll
conditions being satisfied once the running of the self-coupling at
two-loops is included, the constraints imposed from the cosmic
microwave background temperature anisotropies measurements make the
predictions of such a scenario incompatible with the measured value of
the top quark mass.

Finally, running of the gravitational constant and corrections by
considering the more appropriate de\,Sitter, instead of a Minkowski,
background do not improve substantially the realization of a
successful inflationary era~\cite{mmm}.

\vskip 1.truecm

The noncommutative spectral action provides in
addition to the Higgs field, another (massless) scalar
field~\cite{ali-sigma}, denoted by $\sigma$, which is unlike all other
fields in the theory, such as the Higgs field and gauge fields. Note
that $\sigma$ does not exhibit a coupling to the matter sector.

Including this field, the cosmologically relevant terms in the Wick
rotated action read~\cite{ali-sigma}
\be {\cal S}=\int\left[\frac{1}{2\kappa^2} R - \xi_ H R H^2 - \xi_\sigma R
\sigma^2 -\frac{1}{2}(\nabla H)^2 -\frac{1}{2}(\nabla \sigma)^2 -
V(H,\sigma)\right]\ \sqrt{-g}\ d^4x~, \ee
where
\be
V(H,\sigma)=\lambda_H H^4-\mu_H^2H^2+\lambda_\sigma
\sigma^4+\lambda_{ H\sigma}|H|^2\sigma^2~.
\ee
The constants are related to the underlying parameters as
follows:
\begin{align}
\xi_H &=\frac{1}{12}~~~~, &\xi_\sigma
&=\frac{1}{12}~~~~~~~~~~~~~~~~~~~~~~~~~~~~~~~~~~~~\\ \lambda_H
&=\frac{\pi^2\mathfrak{b}}{2f_0\mathfrak{a}^2}~~~~, &\lambda_\sigma &=
\frac{\pi^2\mathfrak{d}}{f_0\mathfrak{c}^2}\\ \mu_H
&=2\Lambda^2\frac{f_2}{f_0}~~~~,
&\lambda_{H\sigma}&=\frac{2\pi^2\mathfrak{e}}{a\mathfrak{c}f_0}~.
\end{align}
In conclusion, neither the $\sigma$ field can lead to a {\sl
  successful} slow-roll inflationary era, if the coupling values are
  conformal~\cite{mmm}.

One should then examine whether the dilaton field, a dynamical field
that can replace the cut-off $\Lambda$, could play the r\^ole of the
inflaton~\footnote{In particular on non-compact spaces where the Dirac
operator has no longer a discrete spectrum.}.  Then the operator
${\cal D}/\Lambda$ is replaced by $e^{-\Phi/2}{\cal D}e^{-\Phi/2}$,
where $\Phi$ stands for the dilaton field~\cite{sdw-coeff}. The action
for the Gravity-Dilaton-Higgs sectors, was then shown to
be~\cite{sdw-coeff}
\beq {\cal S}_{\rm GDH}&=&\int \sqrt{G}\left[
-\frac{1}{2\kappa_0^{2}}R+\frac{1}{2}\left( 1+\frac
{6}{\kappa^{2}_0f^{2}}\right)
G^{\mu\nu}\partial_{\mu}\Phi\partial_{\nu}
 \Phi\right.~\nonumber\\
&&~~~~\left.+G^{\mu\nu}D_{\mu}H^{^{\prime}\ast}D_{\nu
}H^{^{\prime}}-V_{0}\left(
H^{^{\prime}\ast}H^{^{\prime}}\right)\right]d^{4}x~, \eeq
where $G_{\mu\nu}$ is the metric in Einstein frame and $f$ stands for
the dilaton decay constant. The scale $f$ is of the order of the
Planck scale. The dilaton $\Phi$ can be related to a scalar field
$\tilde\sigma$ of dimension one through $\Phi=(1/f)\tilde\sigma$. It
is worth noting that the difference between the above action and the
spectral one is that the latter has a conformal coupling between the
background geometry, in other words the Ricci curvature $R(G)$, and
the Higgs field $H$, which is required in order to get scale invariant
matter couplings. 

Certainly, to investigate whether the dilaton field $\Phi$ could play
the r\^ole of the inflaton, one should first calculate its potential.
\section{Conclusions}
In Connes' formulation of noncommutative geometry, which we have
adopted here, mathematical and physical notions are described in terms
of spectral properties of operators.  By extending the one-to-one
correspondence between spaces and commutative $C^\star$-algebras to
noncommutative algebras, Connes' approach aims at mapping notions of
differential geometry into algebraic terms. The topology of space is
described in terms of the algebras, or equivalently, the properties of
space are encoded in some continuous fields. The model depends
crucially on the choice of the algebra ${\cal A}$, represented on a
Hilbert space ${\cal H}$, and the generalized Dirac operator ${\cal
D}$.  These ${\cal A, H, D}$ form a spectral triple, a fundamental
ingredient of the whole formalism, which contains the information on
the geometry of space-time. The Dirac operator ${\cal D}$ describes
the metric aspects of the model and the behavior of the fundamental
matter fields represented by vectors of the Hilbert space ${\cal H}$.
The fluctuations of the Dirac operator ${\cal D}$ contain the boson
fields, including the mediators of the forces and the Higgs field.

This noncommutative spectral geometry model has been introduced as an
approach to the standard model of particle physics coupled to gravity.
By considering the standard model of strong and electroweak
interactions as a phenomenological model, one tries to retrieve the
noncommutative geometry of space-time.  It turns out that the geometry
can be considered as the product ${\cal M}\times{\cal F}$ of a
four-dimensional smooth compact Riemanian manifold ${\cal M}$ by a
discrete finite noncommutative space ${\cal F}$ composed of just two
points.  The choice of the discrete space is specified by the
symmetries of the Hilbert space in which quarks and leptons are
placed.

The physical picture is straightforward. The left- and right-handed
fermions are placed on two different sheets. The Higgs fields are just
the gauge fields in the discrete dimensions.  The inverse of the
separation between the two sheets can be interpreted as the
electroweak energy scale. It is interesting to remark that this
picture is similar to the the Randall-Sundrum scenario, where a
four-dimensional brane is embedded into a five-dimensional manifold as
a three-dimensional brane placed at $x_5=0$ and $x_5=\pi r_{\rm
comp}$, with $r_{\rm comp}$ the compactification radius.

The choice of a discrete space consisting of only two points can be
(naively) criticized as a simplified approach. However, the two-sheeted
construction has a deeper physical meaning.  The doubling of the
algebra is related to dissipation and the gauge field structure,
required to explain the standard model of particle physics.  Moreover,
by applying 't Hooft's conjecture, stating that loss of information
within completely deterministic dynamics can lead to a quantum
evolution, dissipation may then lead to quantum features. Thus, the
classical construction of noncommutative spectral geometry carries
implicit in the doubling of the algebra the seeds of quantization.

The noncommutative spectral geometry model lives by construction at
very high energy scales. It hence provides a natural framework to
study early universe cosmology. In other words, it motivates a
particular gravitational model which applied to a given cosmological
background can lead to interesting observational consequences.

\vskip1.truecm
%
It is a pleasure to thank the organizers of the Workshop on Non
Commutative Field Theory and Gravity, held in the beautiful island of
Corfu, for inviting me to present this work during a stimulating and
interesting meeting.

\end{document}